\documentclass[aps, prl, reprint, superscriptaddress, nobibnotes]{revtex4-2}
\usepackage{graphicx}
\usepackage{dcolumn}
\usepackage{bm}
\usepackage{booktabs}
\usepackage{mathtools}
\DeclarePairedDelimiter\abs{\lvert}{\rvert}%
\DeclarePairedDelimiter\ave{\langle}{\rangle}%
\usepackage{upgreek}
\usepackage{xcolor}
\usepackage{multirow}

\begin{document}

\preprint{APS/123-QED}
\title{\textbf{Benchmarking Chemically Scalable Machine-Learning Interatomic Potentials for Large-Scale Simulations of Multicomponent Alloys}
}
\author{Fei Shuang}
\thanks{These authors contributed equally to this work.} 
\email{shuangfei@imech.ac.cn}
\affiliation{
State Key Laboratory of Nonlinear Mechanics, Institute of Mechanics, Chinese Academy of Sciences, Beijing, China
}
\affiliation{
Department of Materials Science and Engineering, Faculty of Mechanical Engineering, Delft University of Technology, Mekelweg 2, 2628 CD Delft, the Netherlands
}
\affiliation{
Faculty of Engineering Technology, University of Twente, Drienerlolaan 5, 7522NB Enschede, the Netherlands
}

\author{Penghua Ying}
\thanks{These authors contributed equally to this work.} 
\affiliation{
Laboratory for multiscale mechanics and medical science, SV LAB, School of Aerospace, Xi’an Jiaotong University, Xi’an, Shaanxi, 710049, China
}

\author{Kai Liu}
\affiliation{
Department of Materials Science and Engineering, Faculty of Mechanical Engineering, Delft University of Technology, Mekelweg 2, 2628 CD Delft, the Netherlands
}

\author{Zixiong Wei}
\affiliation{
Department of Materials Science and Engineering, Faculty of Mechanical Engineering, Delft University of Technology, Mekelweg 2, 2628 CD Delft, the Netherlands
}

\author{Fengxian Liu}
\affiliation{
Faculty of Engineering Technology, University of Twente, Drienerlolaan 5, 7522NB Enschede, the Netherlands
}

\author{Zheyong Fan}
\affiliation{
College of Physical Science and Technology, Bohai University, Jinzhou 121013, China
}

\author{Minqiang Jiang}
\affiliation{
State Key Laboratory of Nonlinear Mechanics, Institute of Mechanics, Chinese Academy of Sciences, Beijing, China
}
\affiliation{
School of Engineering Science, University of Chinese Academy of Sciences, Beijing, China
}

\author{Poulumi Dey}
\email{P.Dey@tudelft.nl}
\affiliation{
Department of Materials Science and Engineering, Faculty of Mechanical Engineering, Delft University of Technology, Mekelweg 2, 2628 CD Delft, the Netherlands
}

\date{\today}

\newpage
\begin{abstract}

Machine learning interatomic potentials (MLIPs) with broad chemical flexibility are essential for atomistic simulations of compositionally complex alloys, but their deployment in large-scale molecular dynamics requires a balance among accuracy, efficiency, stability, transferability, and uncertainty quantification.
Here, we benchmark two chemically scalable MLIP frameworks, neuroevolution potential (NEP) and graph atomic cluster expansion (GRACE), for 16 elemental metals and their multicomponent alloys.
GRACE-FS shows higher training efficiency and generally better average accuracy, chemical transferability, and finite-temperature robustness, whereas UNEP-v1 provides substantially higher inference speed and remains competitive in selected stress and large-error metrics.
We further show that chemical transferability is closely linked to high-temperature MD stability in highly multicomponent environments and that ensemble-based uncertainty provides a more reliable error indicator than D-optimality for the heterogeneous systems considered here.
Finally, three-million-atom shock simulations demonstrate that UNEP-v1, combined with ensemble uncertainty, enables uncertainty-aware simulations under extreme dynamic conditions, yielding robust global spall-strength predictions while revealing model sensitivity in local damage pathways.
These results provide practical guidelines for selecting and deploying MLIPs in large-scale simulations of multicomponent alloys.

\begin{description}
\item[Keywords]
Neuroevolution potential, graph atomic cluster expansion potential, metals and alloys
\end{description}
\end{abstract}

\maketitle

\section{\label{introduction}Introduction}
Machine learning interatomic potentials (MLIPs) have transformed atomistic modelling across metals \cite{Lei-benchmark,Ito2024-npj}, alloys \cite{Meng-CrCoNi}, and complex inorganic materials \cite{Meng2025-HEC-MLIP,Meng-FeH,Meng-FeC,Ito2025-CM}. Foundational developments began with Behler--Parrinello-type neural network potentials (NNPs) \cite{Behler2007-NNP-RPL} and Gaussian approximation potentials (GAP) \cite{GAP,Lei-npj}, and were subsequently extended by several influential frameworks, including the spectral neighbor analysis potential (SNAP) based on bispectrum descriptors, moment tensor potentials (MTPs) as a systematically improvable polynomial formalism \cite{Novikov2021-MTP2}, deep potential (DP) for scalable deep-learning-based force fields \cite{DP}, and the atomic cluster expansion (ACE) as a systematically organized body-order expansion \cite{drautz2019atomic-PACE}. More recently, pretrained or universal MLIPs, such as MACE~\cite{Meng2025-HEA-shock}, CHGNet~\cite{deng2023chgnet}, eqV2~\cite{liao2023equiformerv2}, SevenNet~\cite{park_scalable_2024}, and Orb~\cite{neumann2024orb}, have further expanded elemental coverage and improved transferability across broad classes of materials. These advances have substantially improved both accuracy and generality. However, many universal models rely on computationally intensive equivariant graph or message-passing architectures, which can make routine long-timescale molecular dynamics (MD) at very large scales considerably more demanding than with lighter-weight MLIP frameworks. As a result, despite their broad applicability and strong predictive performance, current demonstrations of such models have remained focused predominantly on molecular systems or materials simulations at moderate system sizes, typically up to the order of tens of thousands of atoms \cite{chips-ff,UMA}.

Within this landscape, neuroevolution potential (NEP)~\cite{fan2021neuroevolution} offers a compact GPU-native route to balancing accuracy and throughput. NEP builds radial and angular bases from Chebyshev and Legendre polynomials, couples them to shallow neural networks, and trains with natural evolution strategies in GPUMD~\cite{Xu2025-GPUMD}. The design favors simple kernels, coalesced memory access, and low-latency evaluation, enabling very high atom-step rates on data-center and commodity GPUs. NEP has been used for million-atom heat-transport simulations~\cite{dong2024molecular} and efficient path-integral MD~\cite{ying2025highly} while remaining close to first-principles fidelity. More recently, a universal variant, NEP89, has been introduced for inorganic and organic materials across 89 elements, delivering the same computational efficiency for million-atom simulations~\cite{NEP89}.

Graph atomic cluster expansion (GRACE) is a recently developed MLIP framework tailored for chemically complex and multicomponent systems~\cite{PhysRevX.14.021036-GRACE-FS}.
GRACE extends the ACE formalism by introducing graph-based basis functions that capture semi-local interactions while preserving permutation, translation, and rotation invariances. 
Recent work has trained foundational GRACE models on some of the largest available materials datasets ( MPtrj~\cite{deng2023chgnet}, OMat24~\cite{omat24} and Alexandria ~\cite{Alexandria}), reporting a new Pareto frontier between accuracy and per-atom computational cost~\cite{Lysogorskiy2026-GRACE-FM}. 
In particular, the Finnis--Sinclair-type variant (GRACE-FS) is highlighted for its straightforward parallelization and capability for efficient CPU-only inference, with reported timings indicating favourable scaling in LAMMPS~\cite{LAMMPS}.

This context motivates a more practical question for metals and alloys: which MLIP frameworks can simultaneously provide broad elemental scalability and genuine feasibility for extreme-scale MD? 
In this regard, NEP and GRACE-FS are particularly noteworthy. 
Unlike many current universal MLIPs, which are primarily demonstrated at system sizes up to tens of thousands of atoms, both NEP and GRACE-FS are designed in a way that makes million-atom simulations a realistic target. 
NEP has already shown this capability in large-scale simulations via GPUMD, while GRACE-FS has emerged as a promising alternative with a distinct algorithmic design, favourable scaling, and efficient large-scale inference. 
However, existing universal NEP (NEP89) and GRACE-FS models (GRACE-FS-OAM or GRACE-FS-OMAT) are not necessarily optimized for the specific metallic-alloy chemical space, defect configurations, finite-temperature stability, and high-strain-rate deformation processes considered here.
Moreover, because these universal models differ in training data, model settings, and validation protocols, retraining and evaluating NEP and GRACE-FS within a common benchmark framework is necessary for a controlled comparison.
A direct and protocol-consistent comparison between NEP and GRACE-FS is therefore especially timely. 
Such a comparison is important not only for clarifying their relative strengths in accuracy, efficiency, stability, and extrapolation, but also for identifying what is required for MLIPs to move beyond general-purpose accuracy and toward production-scale simulations of compositionally complex materials.

The scientific payoff is immediate for high-entropy materials, where chemically frustrated disorder and multiscale microstructures demand very large simulation volumes. High-entropy alloys benefit from supercells that capture short-range order, clustering, dislocation networks, and grain boundaries, often at scales beyond thousands of atoms~\cite{zhao2017stacking,chen2021direct,Meng-CrCoNi,cao2022maximum}. High-entropy two-dimensional materials~\cite{nemani2023functional} and high-entropy ceramics~\cite{zhang2019review} similarly require both wide elemental coverage and large-scale simulations to probe phase stability, defect thermodynamics, and transport. Related extreme-scale arenas include additive-manufacturing thermal cycles, shock and impact loading, radiation damage cascades, and nanocrystalline grain growth, where million-atom domains are commonly reported or strongly indicated~\cite{zhou2026probing,zepeda2017probing}. 

Against this backdrop, we present a comparative study of NEP and GRACE-FS trained on a chemically diverse dataset covering 16 elemental metals and their representative alloys. 
We first evaluate their predictive accuracy on training and test sets, together with finite-temperature stability and computational efficiency. 
We then examine the applicability of two uncertainty quantification approaches for these models. 
To assess extrapolation behaviour, we systematically analyse how prediction errors evolve with increasing chemical complexity. 
In addition, both models are validated against a range of physics-grounded properties, including elastic constants, defect energetics, and deformation responses. 
Finally, leveraging its superior computational efficiency, we employ NEP to perform a three-million-atom shock simulation of a high-entropy alloy and analyse the associated uncertainties.

\section{\label{method}Materials and Methods}
\subsection{\label{dataset}Training and test datasets}
The datasets used in this work consist of three components. 
First, the training dataset is taken from Ref.~\cite{song2024-NEP16}, which covers 16 metallic elements (Ag, Al, Au, Cr, Cu, Mg, Mo, Ni, Pb, Pd, Pt, Ta, Ti, V, W, Zr) and includes only configurations of pure metals and binary alloys. 

Second, we follow the testing protocol of Ref.~\cite{song2024-NEP16}, which evaluates model transferability using public multicomponent datasets that extend well beyond the unary and binary training domain. In particular, Song et al. assessed force predictions on three independent datasets containing 3-component, 4-component, and even up-to-13-component structures, and further examined formation energies using ternary Materials Project structures and GNoME structures ranging from 2 to 5 components. These test data therefore provide a substantially broader benchmark for evaluating model accuracy outside the training distribution.

Third, to systematically probe chemical transferability, we construct an additional benchmark dataset of multicomponent alloys covering all compositional orders from 2 to 16 elements. The dataset contains 800 structures in total, spanning the 8 possible levels of compositional complexity. For each order, structures are generated with randomized elemental selections, concentrations, and atomic arrangements, with the goal of covering as many distinct elemental combinations as possible rather than exhaustively enumerating the full combinatorial space. This design provides a systematic benchmark for evaluating how model performance changes as the chemical complexity increases from binary to highly multicomponent alloys. All structures are evaluated using density functional theory (DFT) calculations.

\subsection{\label{dft}DFT calculations}

DFT calculations are performed for all newly generated configurations using the Vienna \textit{ab initio} simulation package (VASP)~\cite{VASP}. 
The exchange--correlation interactions are described by the Perdew--Burke--Ernzerhof (PBE) functional~\cite{pbe}, and electron--ion interactions are treated using the projector-augmented-wave (PAW) method~\cite{paw}. 
An energy convergence criterion of $10^{-6}$ eV is used for electronic self-consistency. 
The plane-wave cutoff energy is set to 520 eV. 
Brillouin zone sampling is performed using Monkhorst--Pack grids generated by VASPKIT~\cite{vaspkit}, with a consistent reciprocal-space density of $2\pi \times 0.03~\mathrm{\AA}^{-1}$.

\subsection{\label{nep}Neuroevolution potential}

The NEP models used in this work are identical to the UNEP-v1 model ensemble reported in Ref.~\cite{song2024-NEP16}. 
All UNEP-v1 models and corresponding reference datasets have been previously deposited in the Zenodo repository \url{https://doi.org/10.5281/zenodo.11533864}~\cite{fan_2024_11533864}. 
Therefore, the UNEP-v1 models were not newly trained in the present work. 
The training cost and generation number reported here follow the original UNEP-v1 training protocol, in which the primary model was trained with a batch size of 10,000 structures for 1,000,000 generations using four A100 GPUs~\cite{song2024-NEP16}. 
Accordingly, the training-time comparison in this work should be interpreted as a wall-time comparison between the published UNEP-v1 training protocol and the GRACE training protocol used here, rather than as a universal convergence-rate comparison between the two frameworks.

The UNEP-v1 models were trained using the NEP4 architecture~\cite{song2024-NEP16} implemented in the GPUMD package~\cite{Xu2025-GPUMD} to represent 16 elemental metals and their alloys. In this framework, the total energy is decomposed into site energies as $U = \sum_{i=1}^N U^i$. Specifically, NEP4 maps a descriptor vector $\mathbf{q}^i$ of a central atom $i$ to its site energy $U^i$ using species-dependent neural network (NN) parameters $\mathbf{w}^I$ (collectively denotes the weight and bias parameters) for each atom type $I$:
\begin{equation}
\label{equation:Ui_nep4}
U^i = \sum_{\mu=1}^{N_\mathrm{neu}}w^{(1)}_{\mu, I}\tanh\left(\sum_{\nu=1}^{N_\mathrm{des}} w^{(0)}_{\mu\nu, I} q^i_{\nu} - b^{(0)}_{\mu, I}\right) - b^{(1)}_I,
\end{equation}
which can be formally expressed as $U^i = \mathcal{N}\left(\mathbf{w}^I; \mathbf{q}^i\right)$. Notably, while using species-dependent parameters increases the total number of trainable parameters, it does not significantly increase the computational cost during MD simulations because it only involves a selection of the correct set of NN parameters for a given atom~\cite{song2024-NEP16}.

The training setup of UNEP-v1 utilized descriptor cutoffs of 6 $\rm \AA$ for radial components and 5 $\rm \AA$ for angular components, with angular terms extending up to five-body interactions. The descriptors $\mathbf{q}^i$ are constructed from radial functions $g_n(r_{ij}) = \sum_{k}c^{IJ}_{nk} f_k(r_{ij})$, where the expansion coefficients $c^{IJ}_{nk}$ are also trainable parameters specific to the atom pairs $(I, J)$.

These descriptors were constructed using 5 radial functions expanded via 9 basis functions for both radial and angular parts. The regression model for each species employed a feedforward neural network with a single hidden layer containing 80 neurons, resulting in a 35-80-1 architecture per element. To accurately handle high-energy short-range interactions, a Ziegler-Biersack-Littmark (ZBL) potential~\cite{Ziegler1985} with a 2 $\rm \AA$ cutoff was incorporated into the potential.

Beyond the primary model, the UNEP-v1 release includes an ensemble of 8 separate models. These were generated by systematically varying the hyperparameters, including the number of radial functions and the number of neurons in the hidden layer (see Ref.~\cite{fan_2024_11533864} for details), within the UNEP-v1 training input to ensure statistical robustness and facilitate uncertainty estimation across the chemical space of the 16 elemental metals and their diverse alloys.

\subsection{\label{grace}Graph atomic cluster expansion potential}

The GRACE potential is an extension of the ACE potential. In the ACE formulation, the energy of the system of $N$ atoms is decomposed into atomic contributions as $E = {\Sigma}_i E_i$, where $i = 1,...,N$. For each contribution, it is represented by a function $\mathcal{F}$ of $P$ atomic properties:
\begin{equation}
    E_i = \mathcal{F}({\varphi}_i^{(1)},...,{\varphi}_i^{(P)}).
\end{equation}
The atomic property has the form:
\begin{equation}
    {\varphi}_i^{(p)} = \sum_{\boldsymbol{\nu}}c_{\boldsymbol{\nu}}^{(p)}\boldsymbol{B}_{i\boldsymbol{\nu}},
\end{equation}
where $c_{\boldsymbol{\nu}}^{(p)}$ is the expansion coefficients and $\boldsymbol{B}_{i\boldsymbol{\nu}}$ is the basis functions. Generally, the function $\mathcal{F}$ is a nonlinear function, for example in the form of generalized Finnis-Sinclair potential
\begin{equation}
    E_i = {\varphi}_i^{(1)} + \sqrt{{\varphi}_i^{(2)}}.
\end{equation}

However, in the GRACE formulation, the decomposition of the system energy is not performed in the beginning. First, the system energy is expressed as $E = E(\boldsymbol{\sigma})$, where $\boldsymbol{\sigma} = (\boldsymbol{r}_{1},\mu_{1},...,\boldsymbol{r}_{N},\mu_{N})$ is the configuration depending on the atomic positions $\boldsymbol{r}_{i}$ and atom properties $\mu_{i}$ such as the species or magnetic moments. After attaching single-particle basis functions to each atom, the cluster basis functions $\Phi_{\alpha\boldsymbol{\nu}}$ are obtained from their products. Thereby, the system energy is represented by the linear combination of the cluster basis functions
\begin{equation}
    E = \sum_{\alpha\boldsymbol{\nu}}J_{\alpha\boldsymbol{\nu}}\Phi_{\alpha\boldsymbol{\nu}}(\boldsymbol{\sigma}),
\end{equation}
where $J_{\alpha\boldsymbol{\nu}}$ is the expansion coefficients. Then, the graph topology of the configuration is considered for the simplifications of the cluster basis functions. Specifically, the clusters in the configuration are either star or tree graphs, where directed edges and nodes correspond to bonds and atoms, respectively. To resemble the decomposition of system energy into atomic contributions, the contribution of the graph-based cluster basis functions is assigned to the root nodes, leading to the introduction of atomic bases in the functional form of single-particle basis functions. Next, the tree graphs are decomposed into star graphs on different layers from the root node. In this way, local ACE potentials could be defined on different layers so that the GRACE potential is evaluated in a recursive approach. Finally, the atomic contributions could be written as
\begin{equation}
    E_i = \sum_{k}\lambda_k^{(\mu_i)}\varphi_{ik}^{(0)},
\end{equation}
where $\varphi_{ik}^{(0)}$ is local ACE potential.

Notably, compared to the ACE potential, the GRACE potential does not impose an explicit limit on the number of chemical elements to be parameterized. This is due to the modification of its radial basis function to include pairwise chemical dependence, resulting in the vanishing of the chemical index into the entries of the expansion coefficients. In addition, GRACE models of different complexities are used in this study, namely GRACE-FS, GRACE-1L, and GRACE-2L. The difference between them is in the range of the message passing. Specifically, the GRACE-FS model is similar to Finnis-Sinclair-type ACE model but with chemical embedding to parameterize unlimited number of elements, while GRACE-1L and GRACE-2L represent one-layer and two-layer models, respectively.

In this study, we employ the GRACE-FS framework to train MLIP models with different levels of complexity, quantified by their parameter counts. Following the \textit{gracemaker} definitions \cite{PhysRevX.14.021036-GRACE-FS}, we consider small, medium, and large variants, denoted as GRACE-FS-S, GRACE-FS-M, and GRACE-FS-L, respectively. To further probe the role of architectural design, we additionally train a more complex graph-based model, GRACE-2L, within the same GRACE framework. A uniform cutoff radius of 6~$\rm \AA$ is adopted for all models. Beyond the primary model, we also train 10 additional GRACE-FS-M models for high-quality uncertainty quantification.

\subsection{\label{uq}Uncertainty quantification}
We use ensemble learning and D-optimality to estimate the prediction errors of our MLIPs on unseen atomic configurations. The ensemble method quantifies uncertainty using the maximum deviation of predictions across multiple models~\cite{ACE-UQ}. Additionally, we use D-optimality with the MaxVol algorithm~\cite{podryabinkin2017active} to compute an extrapolation grade for structures relative to the reference dataset.

Ensemble-based uncertainty quantification is applied to both the UNEP‑v1 and GRACE‑FS models, with 8 models used for each case.
For D-optimality based uncertainty quantification, NEP tool sets \cite{nep_active_link} gives out-of-memory errors on 16-element dataset ($>$100,000 structures). Therefore it is only applied to GRACE-FS models with \texttt{gracemaker}.

Following Ref.~\cite{ACE-UQ}, we define the estimated uncertainties for individual atoms ($U_{F,\mathrm{atom}}$) and entire configurations ($U_{E,\mathrm{cfg}}$, $U_{F,\mathrm{cfg}}$) as the maximum deviations across the ensemble:
\begin{equation}
U_{E,\mathrm{cfg}} = \max_k\abs{E_j^k-\ave{E_j}}\,,
\label{eq:UQ-e-cfg}
\end{equation}
\begin{equation}
U_{F,\mathrm{atom}} = \max_k\abs{\mathbf{F}_i^k-\ave{\mathbf{F}_i}}\,,
\label{eq:UQ-f-atom}
\end{equation}
\begin{equation}
U_{F,\mathrm{cfg}} = \max_{i \in j}(\max_k\abs{\mathbf{F}_i^k-\ave{\mathbf{F}_i}})\,,
\label{eq:UQ-f-cfg}
\end{equation}
where $k = 1, \dots, K$ indices the MLIP models, $E_j^k$ and $\mathbf{F}_i^k$ are the individual model predictions, and $\ave{E_j}$ and $\ave{\mathbf{F}_i}$ are the ensemble averages. We compare these uncertainties to their respective ground-truth DFT errors:
\begin{equation}
e_{E,\mathrm{cfg}} = \abs{E_j^\mathrm{DFT}-\ave{E_j}}\,,
\label{eq:error-e-cfg}
\end{equation}
\begin{equation}
e_{F,\mathrm{atom}} = \abs{\mathbf{F}_i^\mathrm{DFT}-\ave{\mathbf{F}_i}}\,,
\label{eq:error-f-atom}
\end{equation}
\begin{equation}
e_{F,\mathrm{cfg}} = \max_{i \in j}\abs{\mathbf{F}_i^\mathrm{DFT}-\ave{\mathbf{F}_i}}.
\label{eq:error-f-cfg}
\end{equation}


\subsection{\label{MD}Simulation and visualization packages}
LAMMPS~\cite{LAMMPS} and GPUMD~\cite{Xu2025-GPUMD} are used for large-scale MD simulations. To evaluate the computational efficiency of the new GPUMD version, we perform inference speed tests using GPUMD-v5.0. OVITO is employed for the visualization and analysis of atomic structures~\cite{Ovito}.

\section{\label{results}Results}

\subsection{\label{mlip training}Training performance of MLIPs}

\begin{figure}[b]
    \centering
    \includegraphics[width=1\linewidth]{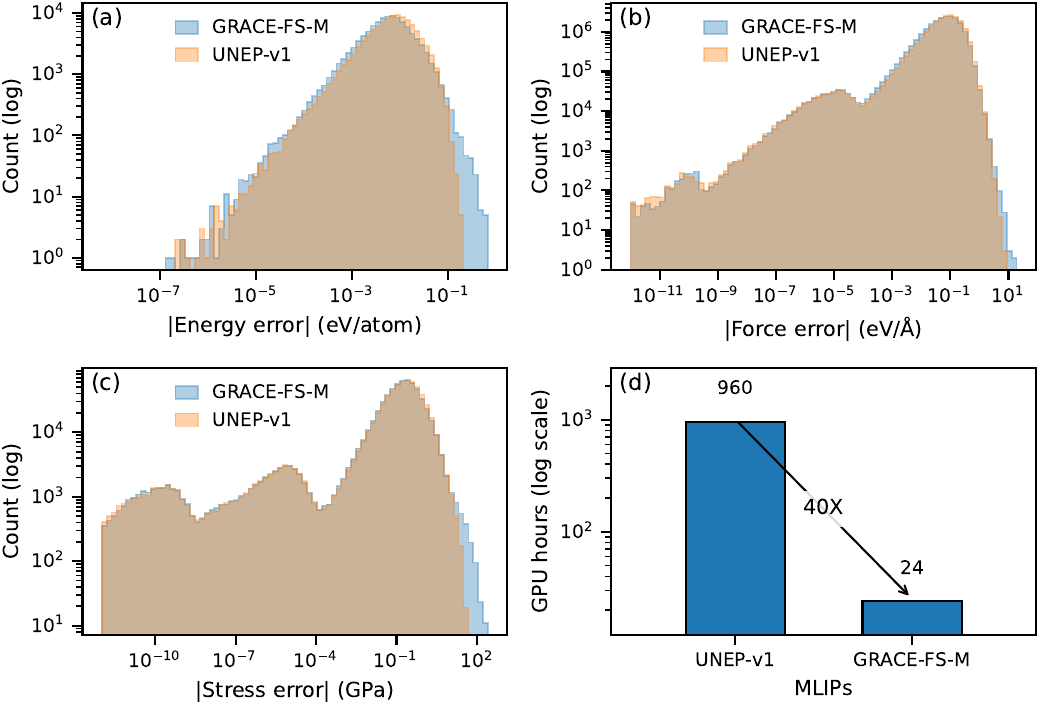}
    \caption{Benchmarking the accuracy and efficiency of UNEP-v1 and GRACE-FS-M potentials. Error distributions for (a) energy, (b) forces, and (c) virial stress are compared alongside (d) the computational cost (wall time) required for training.}
    \label{fig1-accuracy and efficiency}
\end{figure}

We begin by comparing the training performance of UNEP-v1 and GRACE-FS-M in Fig.~\ref{fig1-accuracy and efficiency}. The error distributions for energy, force, and stress in Fig.~\ref{fig1-accuracy and efficiency}(a--c) are broadly similar for the two models. GRACE-FS-M shows a slightly heavier tail in the force distribution and more pronounced high-error tails for energy and stress. In terms of mean absolute error (MAE), GRACE-FS-M yields slightly lower errors for energy (10.42 meV/atom) and force (99.11 meV/\AA) than UNEP-v1 (11.10 meV/atom and 111.08 meV/\AA, respectively), whereas UNEP-v1 gives a lower stress MAE (0.45 GPa versus 0.51 GPa). The root-mean-square errors (RMSEs) show a similar overall trend for force, with GRACE-FS-M again slightly outperforming NEP (159.14 versus 171.90 meV/\AA), but they also reflect the stronger sensitivity of RMSE to the high-error tails: for energy and stress, the RMSEs of GRACE-FS-M (21.02 meV/atom and 2.52 GPa) exceed those of UNEP-v1 (17.13 meV/atom and 1.18 GPa), consistent with the broader tails in the corresponding error distributions. Overall, the two models exhibit comparable training accuracy, with GRACE-FS-M performing slightly better in the average energy and force errors, and UNEP-v1 showing better robustness against large energy and stress outliers.

Training efficiency shows a much sharper contrast. As shown in Fig.~\ref{fig1-accuracy and efficiency}(d), UNEP-v1 required ten days on four A100 GPUs to complete 1{,}000{,}000 generations \cite{song2024-NEP16}, whereas GRACE-FS-M converged within one day on a single A100 GPU in fewer than 2{,}000 epochs. Under the reported training protocols, this corresponds to an approximately 40-fold difference in wall time, highlighting a major practical advantage of GRACE-FS-M for model development and active-learning iteration.

To examine the effect of model complexity, we further consider GRACE-FS-S, GRACE-FS-L, and the two-layer graph-based model GRACE-2L. The numbers of parameters and training errors are listed in Table~\ref{tab1-training-mlip}, and the corresponding test errors are given in Table~\ref{tab2-test-mlip}. The test set from Ref.~\cite{song2024-NEP16} contains pure metals and alloys under diverse deformation states. All GRACE-FS models contain fewer parameters than UNEP-v1, whereas GRACE-2L is substantially larger.

\begin{table*}[ht]
    \setlength{\tabcolsep}{8pt}
    \centering
    \caption{Mean absolute errors (MAE) and root-mean-square errors (RMSE) of energy ($E$), force ($F$), and stress ($S$) for different MLIPs on the training set.}
    \label{tab1-training-mlip}
    \begin{tabular}{lccccc}
        \toprule
        MLIPs & UNEP-v1 & GRACE-FS-S & GRACE-FS-M & GRACE-FS-L & GRACE-2L \\
        \midrule
        \# of Parameters & 70{,}401 & 9{,}312 & 17{,}824 & 28{,}592 & 1{,}289{,}704 \\
        $E_{\mathrm{MAE}}$ (meV/atom) & 11.10 & 11.81 & 10.42 & 9.30 & 3.29 \\
        $E_{\mathrm{RMSE}}$ (meV/atom) & 17.13 & 23.75 & 21.02 & 18.93 & 14.09 \\
        $F_{\mathrm{MAE}}$ (meV/\AA) & 111.08 & 103.85 & 99.11 & 95.95 & 55.15 \\
        $F_{\mathrm{RMSE}}$ (meV/\AA) & 171.90 & 168.59 & 159.14 & 153.73 & 88.15 \\
        $S_{\mathrm{MAE}}$ (GPa) & 0.45 & 0.53 & 0.51 & 0.50 & 0.35 \\
        $S_{\mathrm{RMSE}}$ (GPa) & 1.18 & 2.46 & 2.52 & 2.64 & 4.74 \\
        \bottomrule
    \end{tabular}
\end{table*}

For all models, the energy errors on the test set are higher than those on the training set, indicating a clear generalization gap. This trend is seen in both MAE and RMSE, suggesting not only a shift in the average prediction error but also a deterioration in the larger-error regime on unseen configurations. This suggests that training only on pure metals and binary alloys is insufficient to fully represent the more complex chemical environments in the test set. Within the GRACE-FS series, increasing model size leads to modest reductions in both MAEs and RMSEs for energy and force, with smaller changes for stress. GRACE-FS-M and GRACE-FS-L both outperform UNEP-v1 in the MAEs of energy and force on the training and test sets, whereas NEP yields somewhat lower stress MAEs than all GRACE-FS variants. A similar trend is observed for the force RMSE, for which GRACE-FS-M and GRACE-FS-L remain lower than UNEP-v1. By contrast, for energy and stress, the RMSEs of the GRACE-FS models are not uniformly lower than those of UNEP-v1, indicating that although GRACE-FS improves the average predictive accuracy, UNEP-v1 remains more robust to a subset of larger errors in these quantities.

The ensemble-level statistics further support a more nuanced comparison between UNEP-v1 and GRACE-FS-M, as summarized in Appendix Table~\ref{tab-ensemble-error-uncertainty}. 
For the MAE values, GRACE-FS-M generally exhibits lower ensemble-averaged errors and smaller model-to-model variations than UNEP-v1, especially for energy and force, indicating more consistent average predictive accuracy across the committee models. 
However, the RMSE values do not follow the same uniform trend, since RMSE is more sensitive to a small number of large-error configurations. 
Therefore, while GRACE-FS-M provides more stable average accuracy, UNEP-v1 remains competitive in the large-error regime for some quantities, particularly energy and stress.

GRACE-2L achieves substantially lower MAEs for all three target properties, especially on the test set, indicating the benefit of increased model expressivity and a more flexible graph-based message-passing architecture. Its RMSEs are also generally reduced for energy and force relative to UNEP-v1 and the GRACE-FS models, supporting improved overall transferability, although the stress RMSE remains comparatively high on the training set, suggesting that a small number of large stress errors still persist. However, its parameter count is more than an order of magnitude larger than that of UNEP-v1 and far exceeds those of the GRACE-FS models. Since the primary goal of this work is to compare UNEP-v1 and GRACE-FS as practically deployable MLIPs for chemically complex, large-scale simulations, the most relevant conclusion is that GRACE-FS provides slightly better average accuracy for energy and force than UNEP-v1, whereas UNEP-v1 retains an advantage in stress prediction and in limiting some large energy and stress outliers. 

\begin{table*}[ht]
    \setlength{\tabcolsep}{8pt}
    \centering
    \caption{Mean absolute errors (MAE) and root-mean-square errors (RMSE) of energy ($E$), force ($F$), and stress ($S$) for different MLIPs on the test set from Ref.~\cite{song2024-NEP16}.}
    \label{tab2-test-mlip}
    \begin{tabular}{lccccc}
        \toprule
        MLIPs & UNEP-v1 & GRACE-FS-S & GRACE-FS-M & GRACE-FS-L & GRACE-2L \\
        \midrule
        $E_{\mathrm{MAE}}$ (meV/atom) & 24.44 & 20.07 & 19.89 & 18.94 & 11.54 \\
        $E_{\mathrm{RMSE}}$ (meV/atom) & 57.99 & 62.97 & 55.85 & 68.00 & 64.61 \\
        $F_{\mathrm{MAE}}$ (meV/\AA) & 114.24 & 102.13 & 98.98 & 95.23 & 62.00 \\
        $F_{\mathrm{RMSE}}$ (meV/\AA) & 260.09 & 424.51 & 408.41 & 457.09 & 334.00 \\
        $S_{\mathrm{MAE}}$ (GPa) & 0.67 & 0.65 & 0.62 & 0.61 & 0.37 \\
        $S_{\mathrm{RMSE}}$ (GPa) & 2.13 & 3.07 & 2.47 & 2.86 & 2.18 \\
        \bottomrule
    \end{tabular}
\end{table*}

\subsection{\label{uq}Uncertainty quantification}

\begin{figure*}[ht]
\centering
\includegraphics[width=0.8\linewidth]{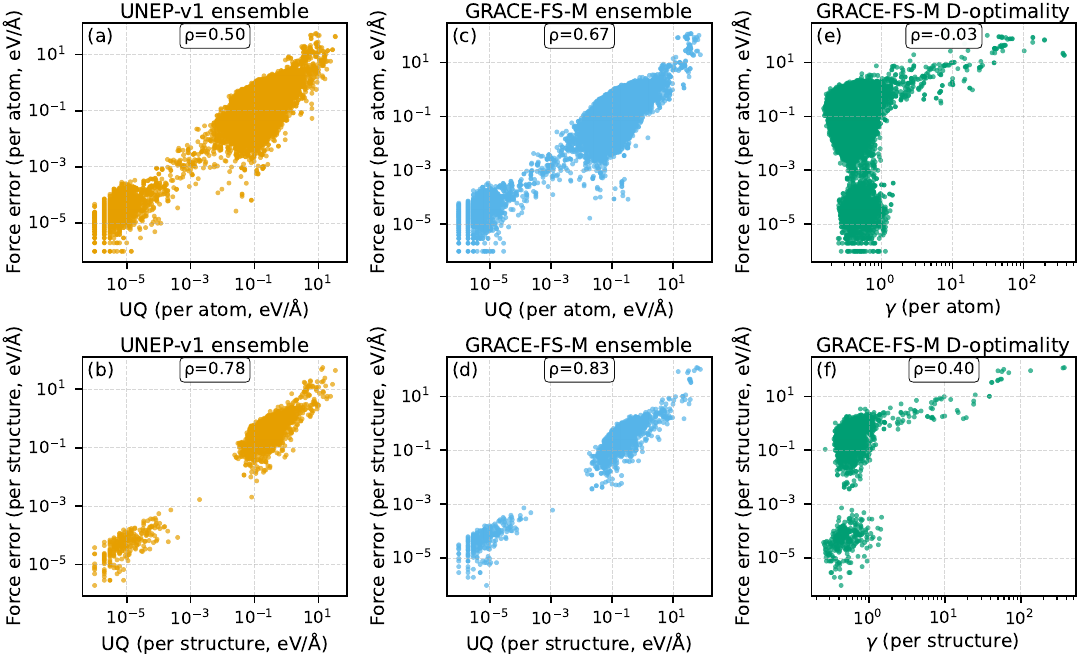}
\caption{Uncertainty quantification for UNEP-v1 and GRACE-FS-M potentials. (a, b) Uncertainty estimates for UNEP-v1 using ensemble learning at the atomic and structural levels, respectively. (c-f) Uncertainty estimates for GRACE-FS-M using (c, d) ensemble learning and (e, f) the D-optimality criterion, each shown at the atomic and structural level.}\label{fig2-uq-error}
\end{figure*}

We next examine the uncertainty of the two MLIPs using ensemble learning and D-optimality approaches. The test dataset from Ref. \cite{song2024-NEP16} is used for the uncertainty quantification (UQ) analysis. The Spearman’s correlation ($\rho$) is used to measure the quality of UQ analysis. Fig. \ref{fig2-uq-error}(a-d) show that using ensemble discrepancies of UNEP-v1 and GRACE-FS-M models can give good estimation of errors for unknown atomic environments. GRACE-FS ensembles exhibit the higher correlation than UNEP-v1 ensembles, potentially because the UNEP-v1 utilize varying model sizes while GRACE-FS-M maintains consistency across models. Notably, for both UNEP-v1 and GRACE-FS-M ensembles, the correlation between force error and UQ is stronger when evaluated on a per-structure basis (Fig. \ref{fig2-uq-error}(b) and (d)) than on a per-atom basis (Fig. \ref{fig2-uq-error}(a) and (c)).

In contrast, the D-optimality results shown in Fig.~\ref{fig2-uq-error}(e, f) exhibit poor correlations between force errors and $\gamma$ values at both the structural and atomic levels.
In particular, the commonly used threshold of $\gamma = 1$ fails to reliably distinguish atoms or structures with low prediction errors, indicating that D-optimality is less reliable for uncertainty quantification in the present heterogeneous 16-element benchmark.
This behavior is consistent with our recent findings on the impact of data heterogeneity on uncertainty quantification~\cite{Shuang2026-ACE-UQ}.
Because the training set spans diverse structural motifs, including severe plastic deformation and free surfaces, as well as 16 elements and their binary alloys, the global descriptor boundary can appear broad while some local regions remain sparsely sampled.
Under such conditions, D-optimality primarily characterizes extrapolation relative to the outer boundary of the sampled descriptor space, but may miss local atomic environments that lie inside this boundary yet are insufficiently represented in the training set.

\subsection{\label{efficiency}Computational efficiency across different scales}
We next evaluate the computational efficiency of the MLIPs, which is critical for enabling extra-large-scale MD simulations.
The performance of UNEP-v1 is benchmarked on both GPU (NVIDIA H100 and A100) and CPU platforms on the Snellius cluster, the national supercomputer of the Netherlands.
CPU calculations are carried out on a dual-socket system equipped with AMD EPYC 9654 processors (192 cores in total).

For a fair comparison with the CPU implementation of GRACE-FS-M, all CPU benchmarks are performed using 192 cores.
This configuration is chosen to match the computational cost of approximately one H100 GPU-hour on the Snellius system, where CPU and GPU nodes are charged at comparable rates.
The resulting performance comparison therefore reflects a cost-normalized evaluation rather than a direct hardware-to-hardware comparison.
In addition, to further disentangle the effects of hardware platform and model architecture, we benchmarked the Kokkos-accelerated implementation of GRACE-FS-M on both CPU and GPU platforms.
Specifically, the CPU results correspond to the double-precision CPU implementation running on 192 CPU cores, whereas the GPU benchmarks were performed using the single-precision Kokkos GPU implementation on NVIDIA A100 and H100 GPUs.
The use of single precision for the GRACE-FS-M GPU benchmarks ensures consistency with the GPUMD implementation used for UNEP-v1, which is also run in single precision.

\begin{figure*}[ht]
\centering
\includegraphics[width=0.8\linewidth]{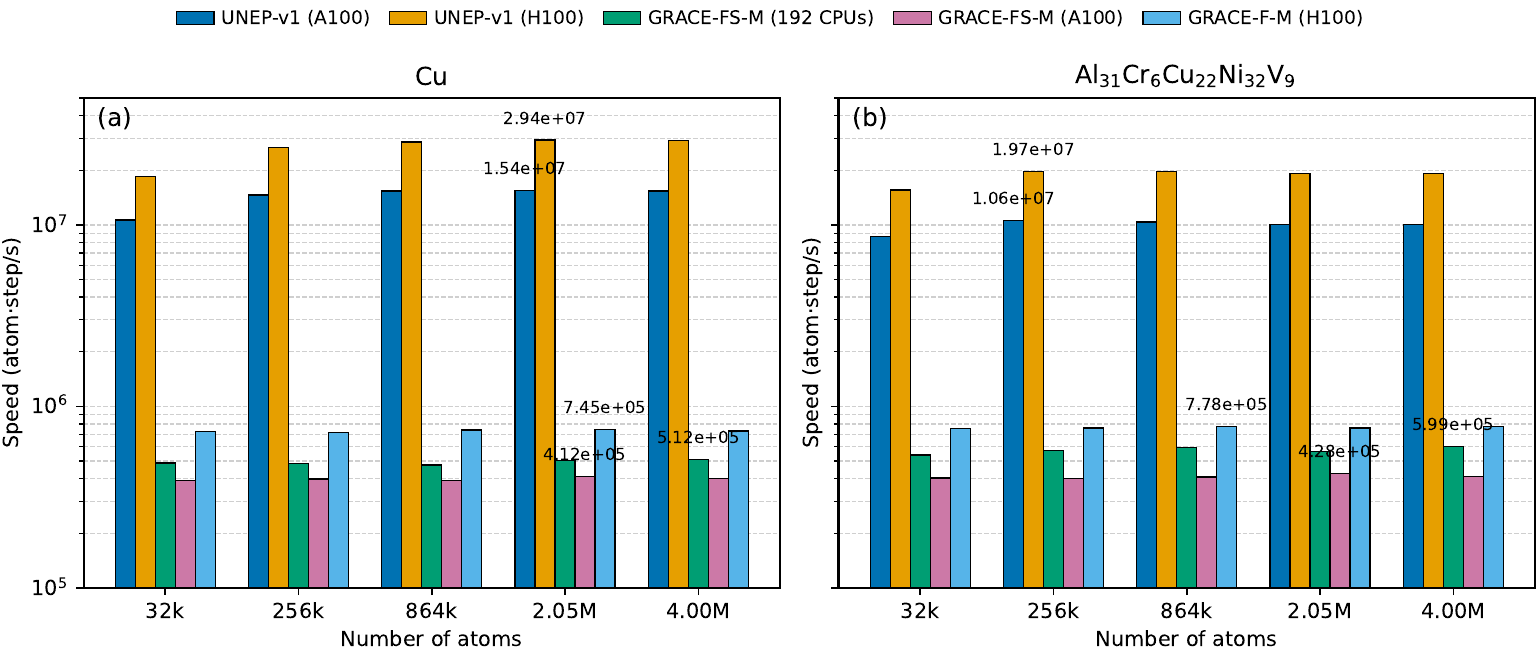}
\caption{Computational speed comparison between UNEP-v1 and GRACE-FS-M for (a) pure Cu and (b) the Al$_{31}$Cr$_{6}$Cu$_{22}$Ni$_{32}$V$_{9}$ HEA. UNEP-v1 is evaluated on A100 and H100 GPUs, whereas GRACE-FS-M is benchmarked using its double-precision CPU implementation on 192 CPU cores and its single-precision GPU implementation on A100 and H100 GPUs.}
\label{fig3-compare-speed}
\end{figure*}

The results, summarized in Fig.~\ref{fig3-compare-speed}, demonstrate a substantial performance advantage for the GPU-accelerated UNEP-v1 over GRACE-FS-M.
For pure Cu, the H100 GPU reaches a peak inference speed of 29.4 million atoms per step per second (Matom$\cdot$step/s) at 2.05 million atoms, compared with 0.503 Matom$\cdot$step/s for the double-precision GRACE-FS-M CPU implementation on 192 CPU cores, corresponding to a cost-normalized speedup of about 58$\times$.
At the same system size, the A100 achieves 15.4 Matom$\cdot$step/s, which is still more than 30$\times$ faster than the CPU reference.
For the chemically more complex Al$_{31}$Cr$_{6}$Cu$_{22}$Ni$_{32}$V$_{9}$ HEA, the H100 attains a peak speed of 19.7 Matom$\cdot$step/s at 256 thousand atoms, whereas the double-precision GRACE-FS-M CPU implementation reaches 0.571 Matom$\cdot$step/s, yielding a cost-normalized speedup of about 34$\times$.
The A100 reaches 10.6 Matom$\cdot$step/s for the same HEA system, maintaining a clear advantage over the CPU-based reference.

We further benchmarked the single-precision Kokkos GPU-accelerated implementation of GRACE-FS-M on A100 and H100 GPUs to disentangle the effects of hardware platform and model architecture.
For pure Cu, GRACE-FS-M reaches only about 0.73--0.74 Matom$\cdot$step/s on H100, while UNEP-v1 reaches 29.2--29.4 Matom$\cdot$step/s at large system sizes, corresponding to an approximately 39--40$\times$ speed advantage even when both models are run on H100 GPUs in single precision.
For the HEA system, GRACE-FS-M reaches about 0.76--0.78 Matom$\cdot$step/s on H100, whereas UNEP-v1 reaches 19.2--19.7 Matom$\cdot$step/s, corresponding to an approximately 25--26$\times$ speed advantage.
These results show that the large performance gap cannot be attributed simply to CPU versus GPU hardware, but primarily reflects differences in model architecture, computational complexity, and implementation efficiency.

The scaling behavior of UNEP-v1 and GRACE-FS-M also differs significantly.
GRACE-FS-M shows nearly system-size-independent throughput over the tested range on both CPU and GPU platforms, suggesting that Kokkos GPU acceleration changes the absolute speed but not the overall scaling trend.
By contrast, UNEP-v1 exhibits a clear increase in GPU throughput with increasing system size and reaches peak performance for systems containing hundreds of thousands to millions of atoms.
This contrast indicates that the favorable scaling of UNEP-v1 is not simply a generic GPU effect, but arises from the combination of its lower computational complexity and efficient GPUMD implementation, which better amortizes GPU parallelization overhead at large system sizes.
This makes UNEP-v1 particularly well suited for extra-large-scale molecular dynamics simulations.

A direct comparison between GPUs further shows that H100 consistently outperforms A100 for UNEP-v1, with an advantage of about 1.74--1.91$\times$ for Cu and 1.81--1.99$\times$ for the HEA, depending on system size.
Given that the H100 is only about 1.5$\times$ more expensive than the A100 on the Snellius cluster, it also delivers a superior performance-to-cost ratio for UNEP-v1.
Therefore, the combination of UNEP-v1 accelerated on an H100 GPU emerges as the most powerful and cost-effective computational pathway for extra-large-scale MD simulations of complex material systems in this work.

\subsection{\label{stability}MD Simulation stability}

To ensure a consistent stability comparison between GRACE-FS-M and UNEP-v1, all MD stability tests were performed using double-precision LAMMPS calculations for both potentials. The neighbor-list settings were kept identical, with \texttt{neighbor 2.0 bin} and \texttt{neigh\_modify every 1 delay 0 check yes}, to avoid numerical differences arising from different neighbor-list update protocols. For each potential, six ensemble models were evaluated, and the maximum energy drift during the NVE stage was summarized using the ensemble mean, minimum, and maximum values.

We evaluate the MD stability of the two MLIPs at different temperatures by adapting the procedure from Ref.~\cite{MSMSE-MD-stability}. For each system, we first perform a 10 ps NVT simulation at the target temperature, followed by a 100 ps NVE simulation to monitor the total energy drift. A non-zero energy drift reflects numerical errors of the potential and its implementation, since the total energy should be conserved in an ideal NVE ensemble. In addition to energy conservation, we also examine whether the simulated structures retain their structural integrity after finite-temperature MD.

\begin{figure}[b]
\centering
\includegraphics[width=1\linewidth]{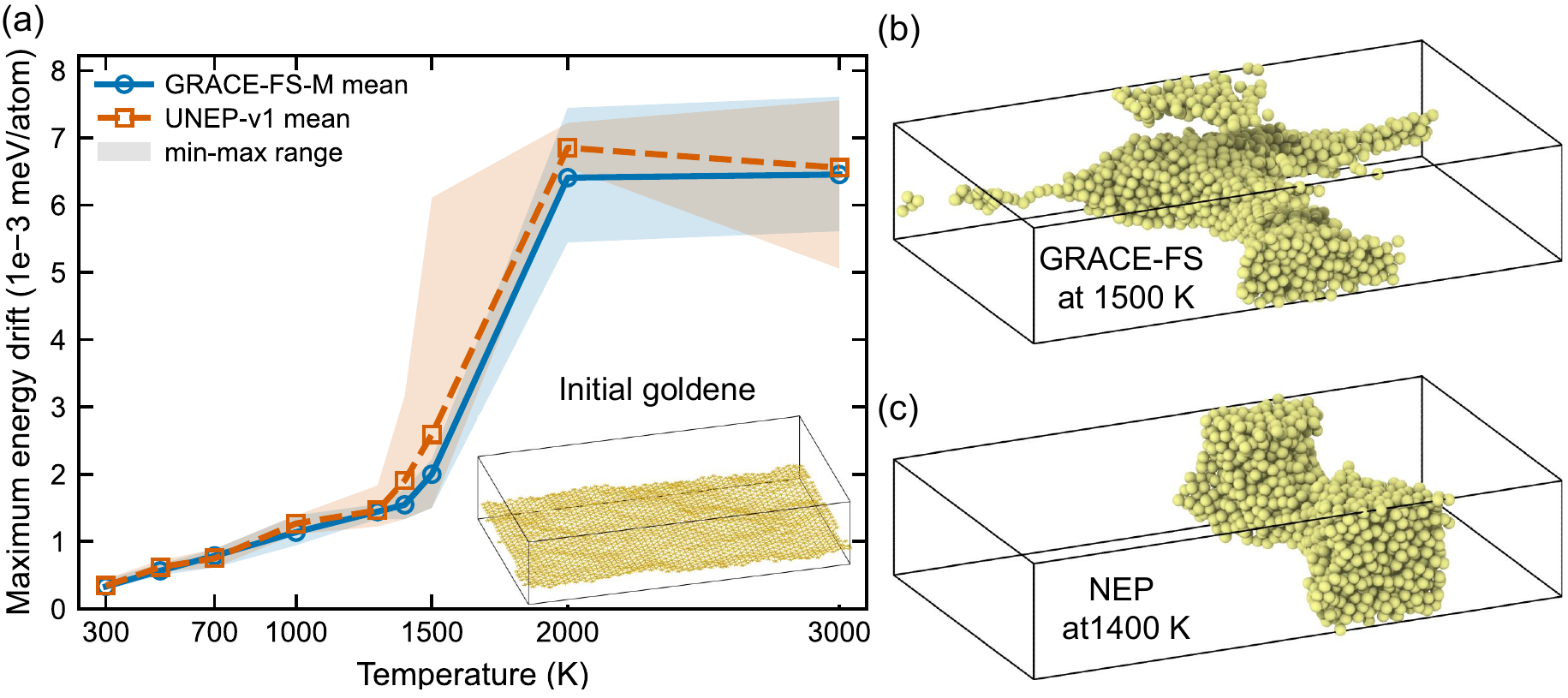}
\caption{Thermal stability of monolayer goldene assessed by MD simulations. (a) Maximum energy drift as a function of temperature. (b, c) Atomic snapshots after finite-temperature simulations, showing the different structural responses predicted by the two MLIPs at high temperature.}
\label{fig4-goldene}
\end{figure}

The results for monolayer goldene are shown in Fig.~\ref{fig4-goldene}. For both MLIPs, the ensemble-averaged maximum energy drift increases gradually from 300 K to approximately 1300 K, indicating comparable energy-conservation behavior at low and moderate temperatures. At higher temperatures, however, UNEP-v1 shows a sharper increase in the maximum drift, reaching $6.11\times10^{-3}$ meV/atom at 1500 K, whereas GRACE-FS-M remains below $2.22\times10^{-3}$ meV/atom. This difference is accompanied by different structural responses. The goldene monolayer starts to lose its structural integrity at 1400 K in the UNEP-v1 simulations, while GRACE-FS-M preserves the monolayer up to 1400 K and collapses only at 1500 K. Thus, the abrupt increase in the UNEP-v1 energy drift coincides with the onset of structural instability.

Previous DFT-based AIMD simulations reported that monolayer goldene remains structurally stable at 1400 K over a 10 ps simulation window~\cite{DFT-gold}. Although this AIMD reference is much shorter than the present 100 ps NVE stability test, it suggests that the GRACE-FS-M prediction is more consistent with the available first-principles evidence in this temperature range. Nevertheless, without longer-time finite-temperature DFT simulations or experimental reference data near 1400--1500 K, we do not assign definitive physical correctness to either prediction. Instead, this result should be interpreted as a model-dependent difference in predicted finite-temperature structural stability.

\begin{figure}[b]
\centering
\includegraphics[width=1\linewidth]{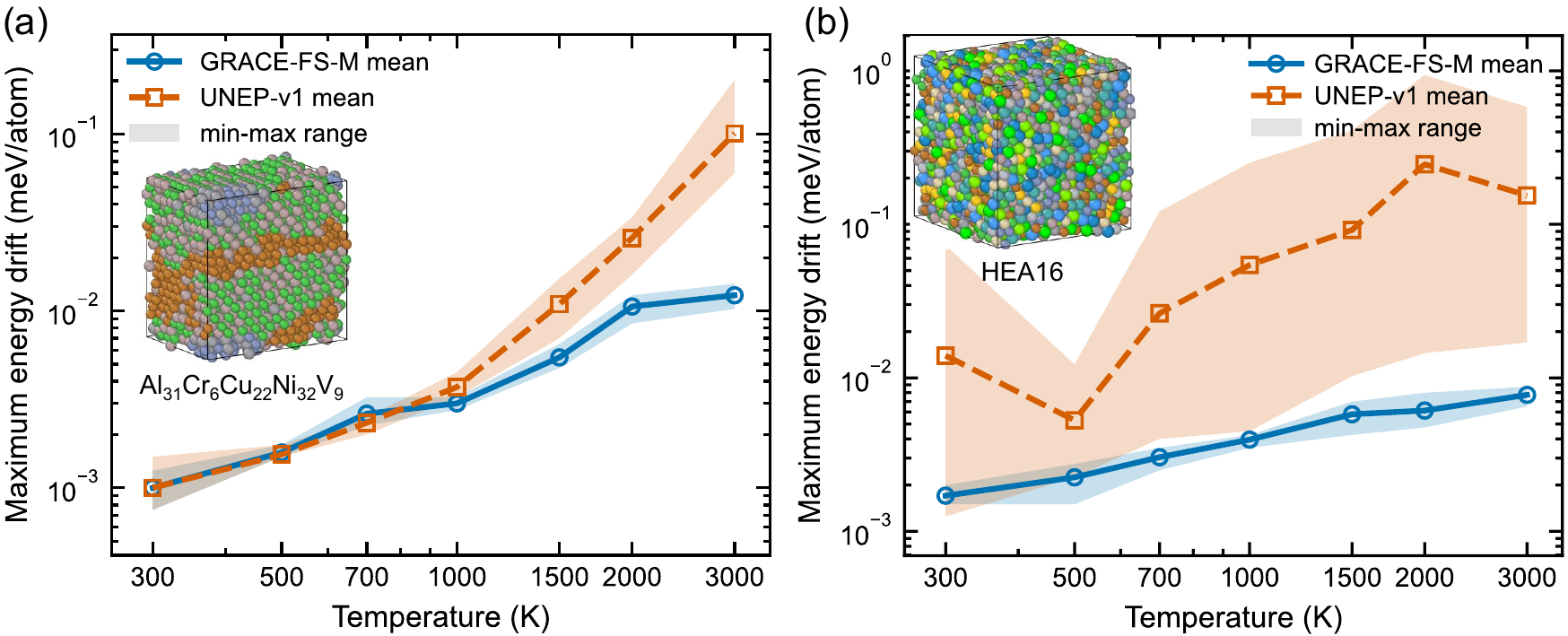}
\caption{Thermal stability of multicomponent alloys assessed by MD simulations. The maximum energy drift is plotted for (a) the Al$_{31}$Cr$_{6}$Cu$_{22}$Ni$_{32}$V$_{9}$ alloy with chemical short-range order (CSRO) and (b) a random 16-element alloy across a range of temperatures.}
\label{fig5-hea}
\end{figure}

We next assess the two MLIPs on more complex multicomponent alloys. The first case is the quinary Al$_{31}$Cr$_{6}$Cu$_{22}$Ni$_{32}$V$_{9}$ alloy with chemical short-range order (CSRO), which was previously studied in Ref.~\cite{song2024-NEP16}. Using this CSRO-containing structure as the initial configuration, we evaluate the energy conservation of both MLIPs over the same temperature range. As shown in Fig.~\ref{fig5-hea}(a), GRACE-FS-M exhibits a smooth and moderate increase in the ensemble-averaged maximum energy drift, from $1.00\times10^{-3}$ meV/atom at 300 K to $1.23\times10^{-2}$ meV/atom at 3000 K. In contrast, UNEP-v1 remains reasonably stable at low and moderate temperatures but shows a much stronger increase at high temperature, with the ensemble-averaged drift reaching $1.01\times10^{-1}$ meV/atom and the maximum value reaching $2.02\times10^{-1}$ meV/atom at 3000 K. Thus, for the quinary alloy, UNEP-v1 exhibits a pronounced loss of energy conservation under extreme-temperature conditions, whereas GRACE-FS-M remains comparatively stable.

The contrast becomes stronger in the random 16-element alloy, as shown in Fig.~\ref{fig5-hea}(b). GRACE-FS-M maintains small and smoothly varying energy drift across the entire temperature range, with the ensemble-averaged drift remaining below $7.75\times10^{-3}$ meV/atom even at 3000 K. By contrast, UNEP-v1 shows much larger drift values and broader model-to-model variation in the 16-element alloy, with the ensemble-averaged drift reaching $2.46\times10^{-1}$ meV/atom at 2000 K and the maximum drift approaching $9.43\times10^{-1}$ meV/atom. These results indicate that the high-temperature instability of UNEP-v1 is strongly amplified by increasing chemical complexity, whereas GRACE-FS-M remains robust from the quinary alloy to the 16-element alloy.

A key observation from these multicomponent systems is therefore not simply that high temperature increases energy drift, but that the influence of elemental complexity is highly model-dependent. GRACE-FS-M maintains small and smoothly varying energy drift in both the five-component and 16-component alloys, indicating good numerical robustness as chemical complexity increases. In contrast, UNEP-v1 becomes much less stable in the 16-element alloy, exhibiting substantially larger ensemble-averaged drift and maximum drift values than in the quinary alloy. This difference is likely associated with their different levels of chemical extrapolation capability, which will be examined in more detail in the next section. These ensemble results suggest that GRACE-FS-M is more robust for high-temperature MD simulations of multicomponent alloys, whereas UNEP-v1 should be used with caution under chemically complex and thermally demanding conditions.

\subsection{\label{transferability}Transferability of MLIPs and augmented GRACE-FS}

The applicability of MLIPs to multicomponent alloys remains a critical open question, as current models are trained exclusively on unary and binary systems. To systematically assess and compare the accuracy of leading MLIPs, we introduce a novel benchmark dataset designed for maximum compositional diversity. This dataset comprises eight distinct systems, ranging from 2 to 16 elements. For each system size, we generated 100 random configurations with varying elemental compositions and atomic perturbations. The performance of each potential is evaluated by comparing its predictions against high-fidelity, single-point DFT calculations. 

\begin{figure}[b]
\centering
\includegraphics[width=1\linewidth]{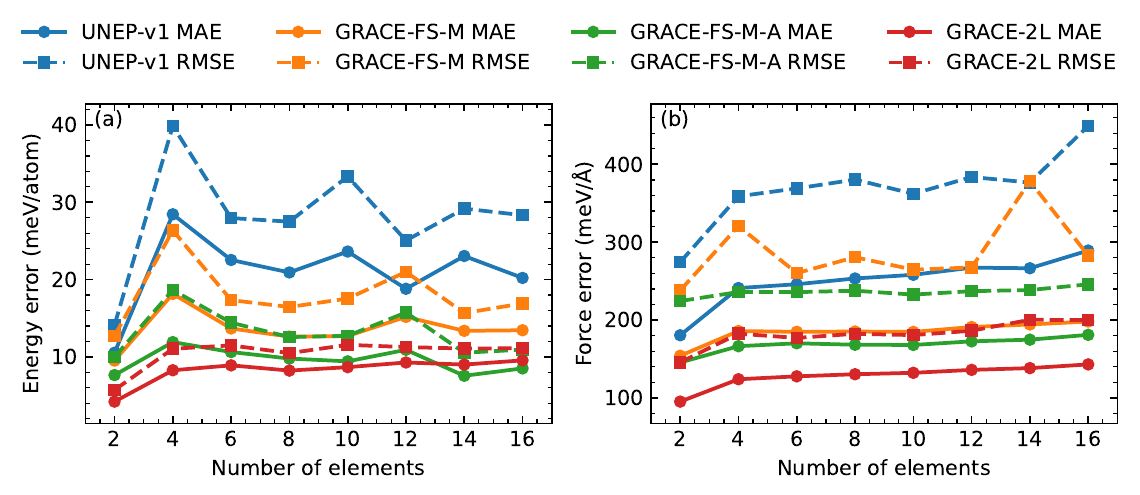}
\caption{Assessing the chemical transferability of different MLIPs. Prediction errors for (a) energy and (b) forces (reported as MAE and RMSE) are evaluated on additional DFT datasets of multi-element systems, ranging from 2 to 16 components.}\label{fig6-transferability}
\end{figure}

Fig.~\ref{fig6-transferability} presents a comparative analysis of the UNEP-v1, GRACE-FS-M, and GRACE-2L models, revealing a clear performance hierarchy: GRACE-2L is the most accurate, followed by GRACE-FS-M, with UNEP-v1 exhibiting the highest errors. 
This result is striking because all models are trained on datasets containing only unary and binary systems, yet they generalize to multicomponent environments with substantially different accuracy. 
This immediately suggests a strong dependence on model architecture. 
To quantitatively investigate this architectural dependence, we trained an augmented model, GRACE-FS-M-A, by supplementing the original training data with our newly generated multicomponent structures. 
These newly generated datasets span configurations ranging from 2 to 16 elements, reflecting increasing chemical complexity. 
The results show a clear trend. 
While data augmentation reduces the error of the baseline GRACE-FS-M, its performance remains inferior to GRACE-2L. 
This suggests that the limitation of GRACE-FS-M is not simply due to a lack of data, but is also related to its architectural design, which constrains its ability to capture complex multi-element interactions. 
Therefore, the superior performance of GRACE-2L, achieved without any multicomponent training data, provides an important insight: for the prediction of complex multicomponent systems, a more expressive model architecture is important for high-fidelity extrapolation and can be more effective than targeted data expansion alone. 

This chemical-transferability trend is also consistent with the MD stability results in Fig.~\ref{fig5-hea}(b), where UNEP-v1 shows much larger energy drift than GRACE-FS-M in the 16-element alloy.
This suggests that limited chemical extrapolation capability may affect not only static prediction accuracy, but also the robustness of finite-temperature MD simulations in highly multicomponent environments.
A plausible origin of the different chemical extrapolation behavior is the distinct way in which chemical species are encoded and information is shared across elemental combinations in the two architectures.
In UNEP-v1, the species-dependent neural-network parameters and pair-dependent descriptor coefficients may make the model more sensitive to chemical combinations that are absent from the unary and binary training data.
By contrast, GRACE-FS-M incorporates chemical embedding within an ACE-like Finnis--Sinclair form, which may enable smoother information sharing across different elemental pairs and multicomponent local environments.
This architectural difference may partly explain the stronger chemical transferability of GRACE-FS-M observed here, although further controlled studies are needed to isolate its precise origin and to guide future improvements of NEP models.

\subsection{\label{validation}Validation of MLIPs for key properties}

\begin{figure}[b]
\centering
\includegraphics[width=1\linewidth]{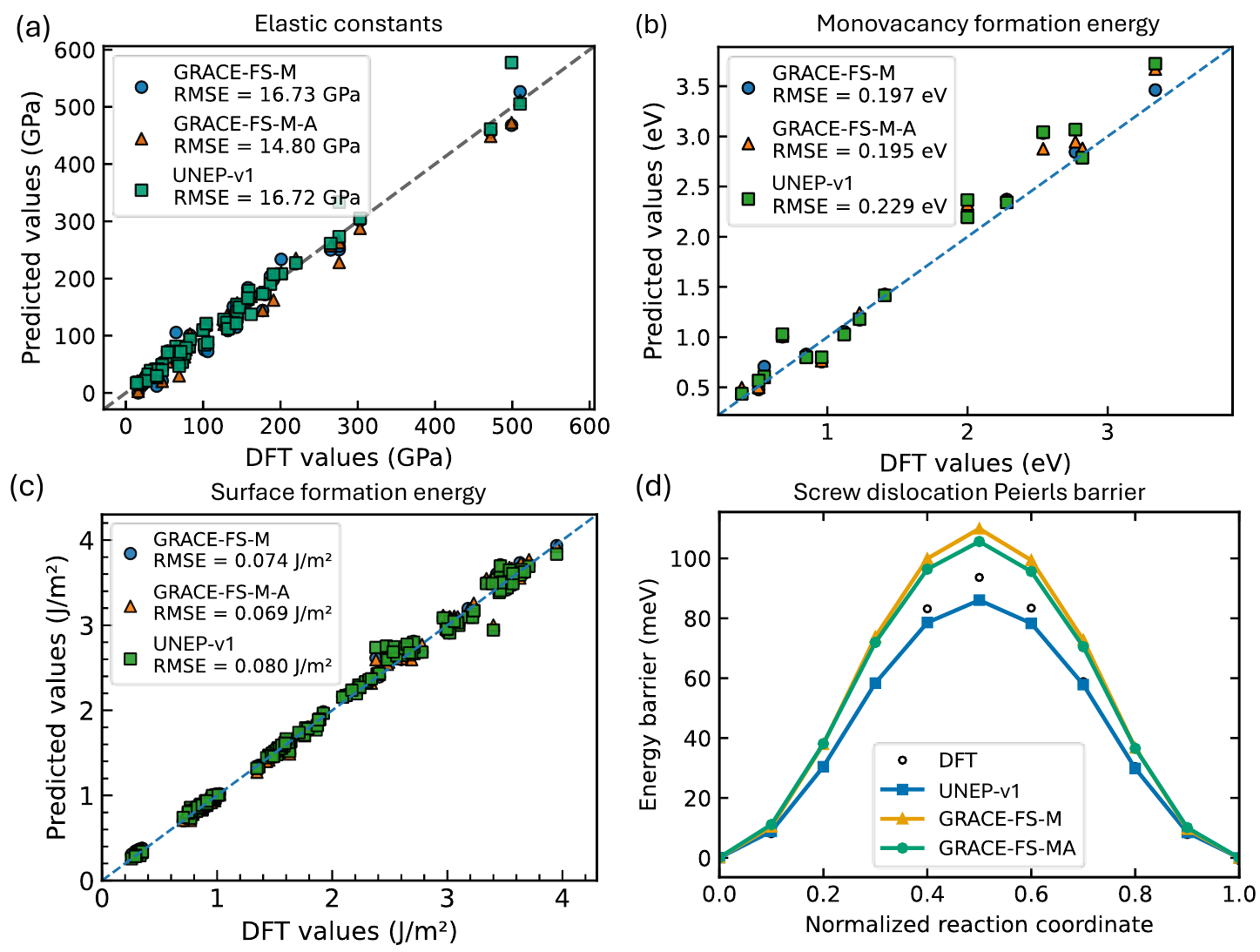}
\caption{Assessment of MLIP accuracy in predicting fundamental mechanical properties. The MLIP-predicted (a) elastic constants, (b) vacancy formation energy, (c) surface energy, and (d) screw dislocation Peierls barrier are plotted against their respective DFT values.}\label{fig7-test-property}
\end{figure}

We next evaluate the accuracy of the MLIPs in predicting key mechanical properties of pure metals, with all reference DFT data obtained from Ref. \cite{song2024-NEP16}. Figure \ref{fig7-test-property}(a) presents a parity plot of the elastic constants for 16 metals. All MLIPs capture the general trends well, with GRACE-FS-M and UNEP-v1 showing comparable errors. The augmented GRACE-FS-M-A model, however, demonstrates a marked reduction in error. For more localized defects, the advantages of the GRACE models become more pronounced. As shown in Fig. \ref{fig7-test-property}(b, c), GRACE-FS-M provides a more accurate prediction of both the monovacancy formation energy and surface energy than UNEP-v1. Furthermore, data augmentation yields consistent improvements, with GRACE-FS-M-A achieving the highest accuracy. Finally, we probe the complex energy landscape of a screw dislocation in BCC tungsten (W) (Fig. \ref{fig7-test-property}(d)). While the UNEP-v1 model underestimates the Peierls barrier, the GRACE-FS models slightly overestimate it; critically, all models yield results close to the DFT reference.

These results collectively demonstrate two key findings: first, the GRACE-FS-M architecture inherently provides superior accuracy for mechanical property prediction, particularly for defect properties, compared to UNEP-v1. Second, the systematic use of data augmentation (GRACE-FS-M-A) consistently enhances accuracy, even for properties not explicitly included in the augmented dataset. This suggests that the augmented data may improve the regularity and coverage of the learned potential energy surface, although the precise origin of this improvement requires further controlled ablation.

\begin{figure}[b]
\centering
\includegraphics[width=1\linewidth]{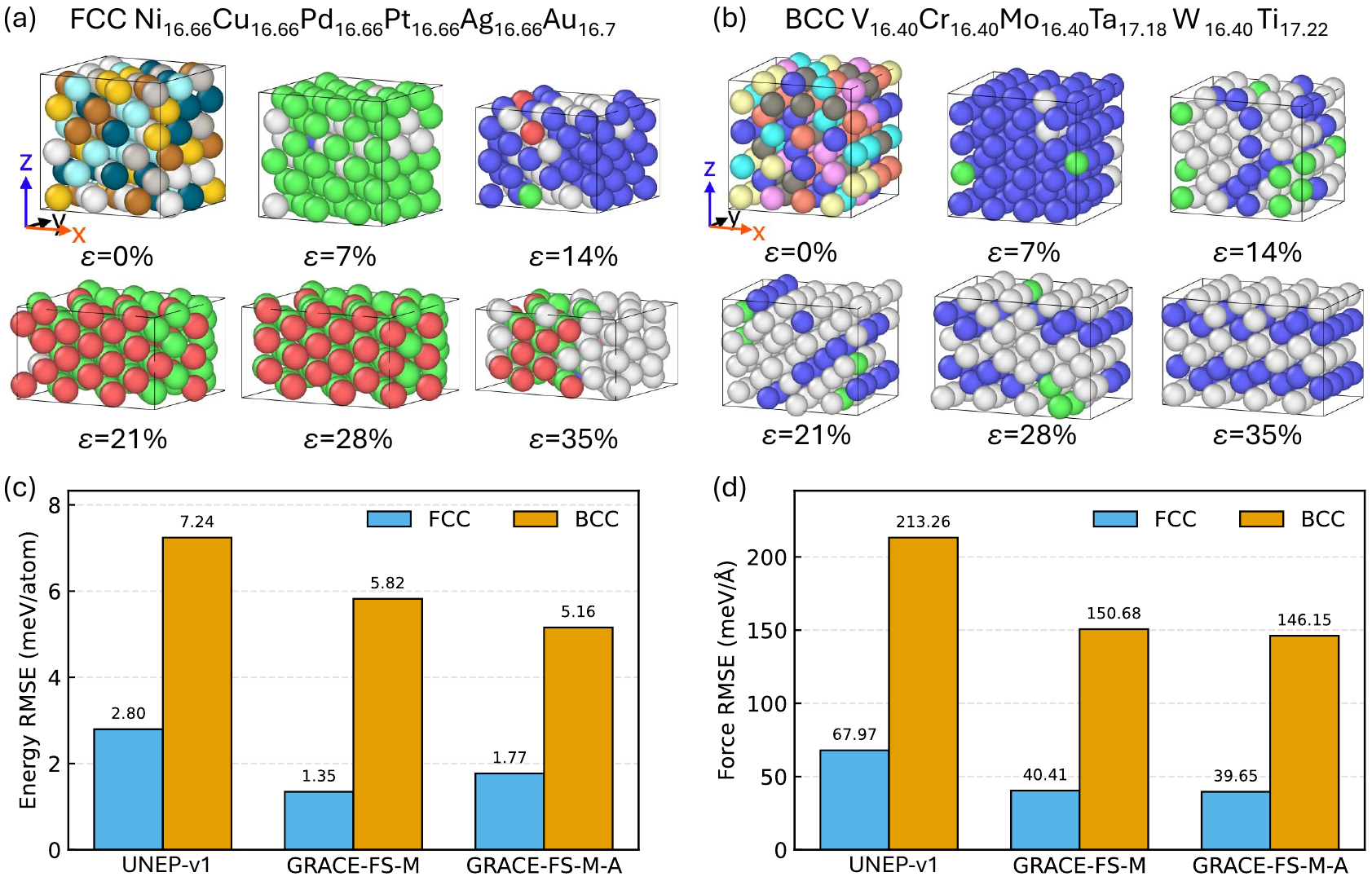}
\caption{Performance evaluation of MLIPs for simulating the tensile testing of HEAs. (a, b) Snapshots of defect structures (e.g., phase transitions, stacking faults) during deformation for FCC and BCC crystal structures, respectively. (c, d) The accuracy of the MLIPs is quantified by RMSE for (c) energy and (d) force predictions compared to reference DFT calculations.}\label{fig8-tensile-test}
\end{figure}

A key validation of the MLIPs involves their application to the finite-temperature deformation of complex, multicomponent systems. For this purpose, we subject two representative high-entropy alloys (HEAs), a FCC Ni-Cu-Pd-Pt-Ag-Au alloy and a BCC V-Cr-Mo-Ta-W-Ti alloy, to tensile loading at a strain rate of 10$^9$ s$^{-1}$ up to 35$\%$ strain. The resultant defect evolution, analyzed via common neighbor analysis (Fig. \ref{fig8-tensile-test}(a, b)), reveals distinct, complex deformation mechanisms. The FCC HEA develops local BCC- and HCP-like environments, accompanied by the formation of unidentified defects. In contrast, the BCC HEA experiences severe plastic deformation, generating a high density of defects and forming systemic boundaries at the maximum strain. The ability to capture such intricate behavior provides a stringent validation for the MLIPs. As shown in Fig. \ref{fig8-tensile-test}(c, d), the GRACE-FS models achieve significantly lower energy and force errors than UNEP-v1 for both HEAs. Furthermore, all potentials exhibit higher accuracy in the FCC structure than in the BCC structure. These results collectively demonstrate that the GRACE-FS architecture provides a more robust and reliable framework for simulating the mechanical response of chemically complex systems, with a particular advantage in capturing the challenging deformation pathways of BCC alloys.

\subsection{\label{shock}Shock simulations and uncertainty analysis}

\begin{figure}[b]
\centering
\includegraphics[width=1\linewidth]{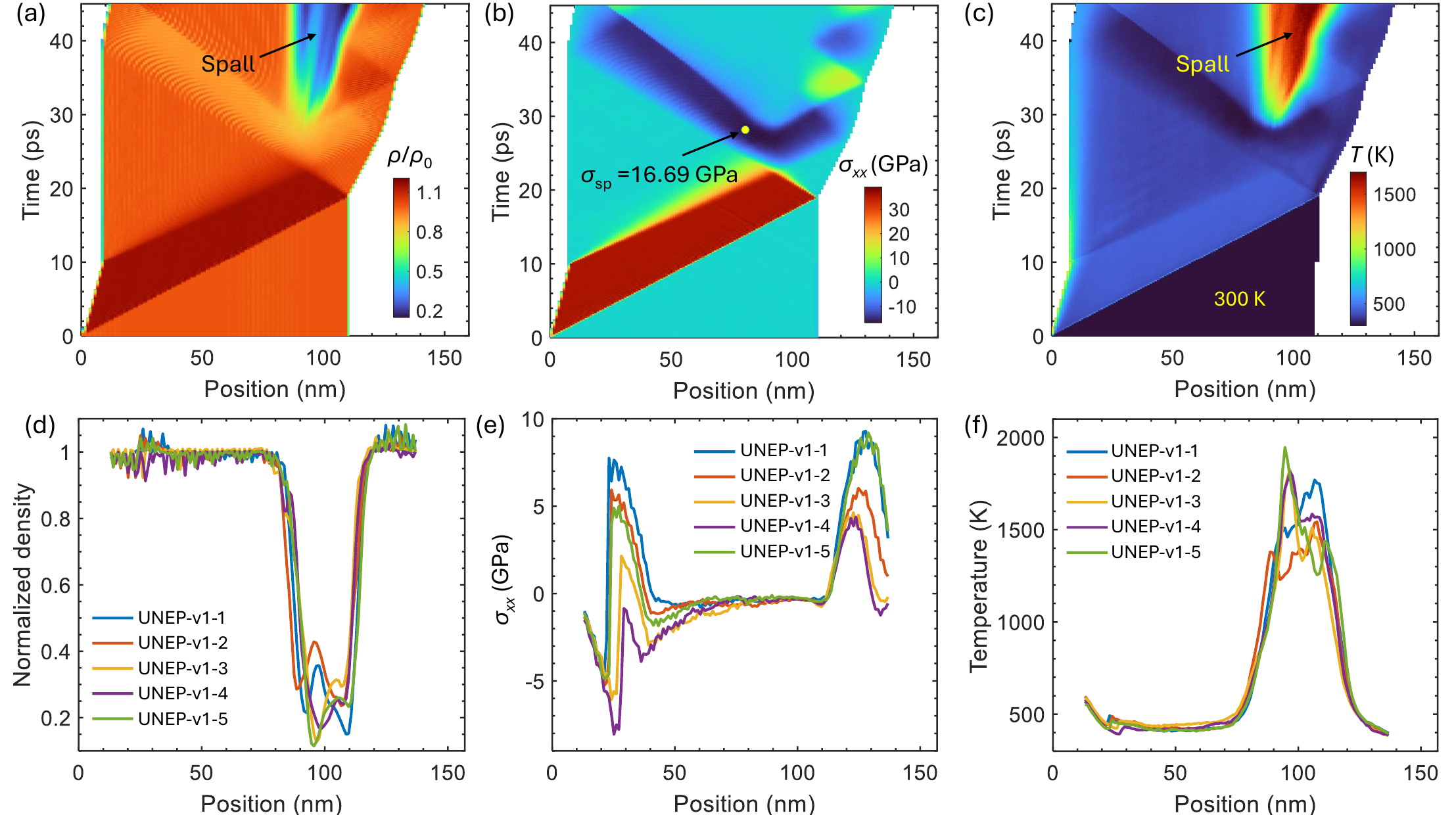}
\caption{Shock-induced spallation in Al$_{10}$Cr$_{10}$Cu$_{35}$Ni$_{35}$V$_{10}$. Temporal evolution of (a) density, (b) stress, and (c) temperature from the primary UNEP-v1 model. The arrows indicate the spallation, and the calculated spall strength ($\sigma_\mathrm{sp}$) is marked in (b). Corresponding spatial profiles of (d) density, (e) stress, and (f) temperature at 45 ps, comparing results across the five-model UNEP-v1 ensemble.}\label{fig9-shock-analysis}
\end{figure}

We next employ the UNEP-v1 ensemble to investigate the shock response of the high-entropy alloy Al$_{10}$Cr$_{10}$Cu$_{35}$Ni$_{35}$V$_{10}$ via non-equilibrium MD (NEMD) simulations. The simulation cell, containing 3 million atoms with dimensions of 18 × 18 × 110 nm$^3$, is subjected to an impact velocity of 0.8 km/s along the [100] crystallographic direction. The shock loading is applied for 10.0 ps using the “ensemble wall\_mirror” method in the GPUMD package \cite{Pan2024-GPUMD-shock}, followed by a 35.0 ps relaxation period to observe subsequent deformation and failure mechanisms. Owing to their high computational demand, GRACE-FS potentials are deemed unsuitable for such application.

Figure \ref{fig9-shock-analysis}(a-c) presents the evolution of density, stress, and temperature during the shock process, as simulated by the primary UNEP-v1 model from Ref. \cite{song2024-NEP16}. The UNEP-v1 model properly captures the dynamic response of these quantities throughout both the shock compression and release stages, including wave propagation, reflection, and interaction. A spall fracture is initiated at approximately 30 ps, marked by a sharp drop in density (Fig. \ref{fig9-shock-analysis}(a)) and a concurrent temperature increase (Fig. \ref{fig9-shock-analysis}(c)). The spall strength, defined as the maximum tensile stress experienced by the material, is determined to be 16.69 GPa for this primary model. To assess model uncertainty, the spall strength was calculated using four additional UNEP-v1 models from the ensemble, yielding values of 17.07, 17.70, 17.47, and 16.76 GPa. The standard deviation across all five models is 0.39 GPa, representing an uncertainty of only $\sim$2.3$\%$ relative to the mean value of 17.14 GPa. This low ensemble spread supports the robustness of the predicted spall strength under the present loading condition. In contrast, Fig. \ref{fig9-shock-analysis}(d-f) shows the spatial distributions of these quantities at the simulation end time (45 ps) for all five UNEP-v1 models. These snapshots reveal non-negligible deviations in critical regions, particularly in the density and temperature profiles within the spall zone.

\begin{figure}[b]
\centering
\includegraphics[width=1\linewidth]{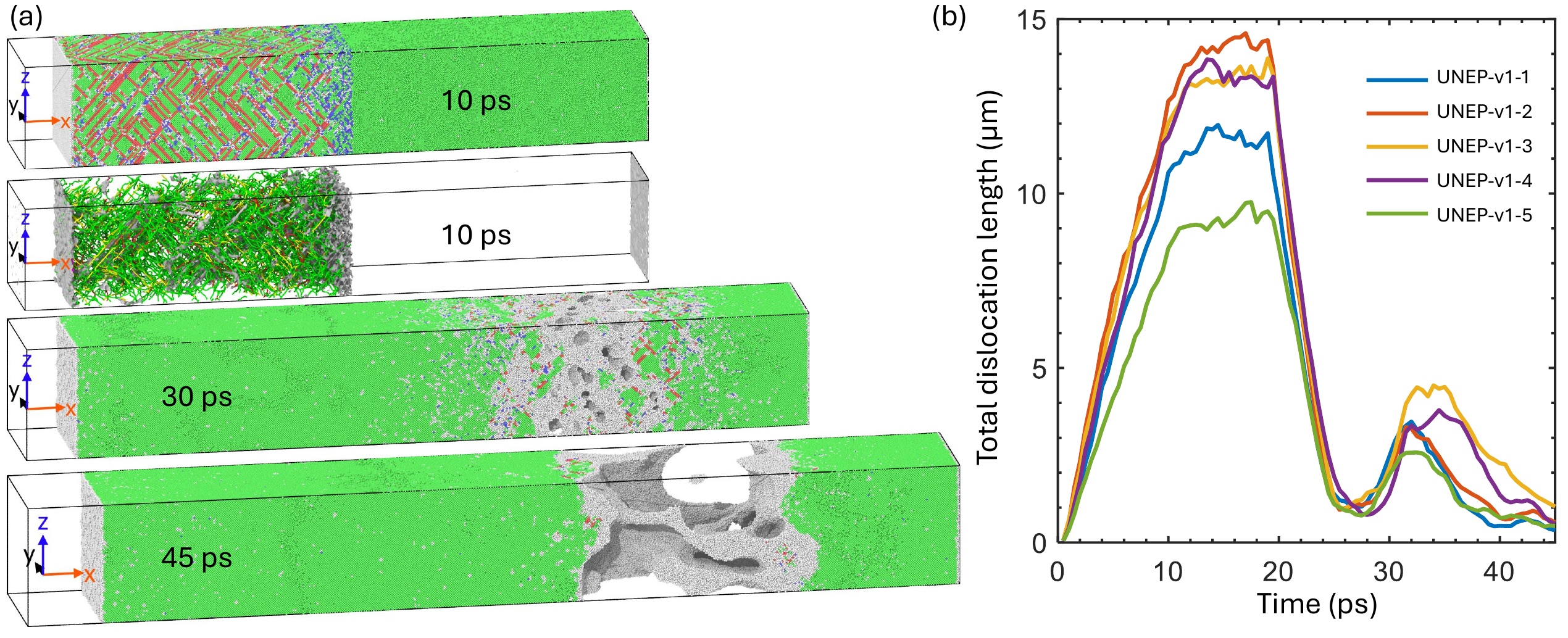}
\caption{Microstructural evolution during shock simulation. (a) Temporal evolution of defect structures. The first panel shows dislocation lines extracted using the Dislocation Extraction Algorithm (DXA), whereas the subsequent panels show defects identified by Common Neighbor Analysis (CNA). (b) Temporal evolution of the total dislocation length for five different UNEP-v1 models.}\label{fig10-shock-config}
\end{figure}

We next analyze the microstructure evolution and failure mechanisms. 
Figure~\ref{fig10-shock-config}(a) presents the sequence from the main UNEP-v1 model. 
By the end of the shock stage (10 ps), a dense dislocation network has formed. 
Subsequently, void nucleation occurs at the right end of the sample by 30 ps, leading to complete fracture into two parts by 45 ps. 
The temporal evolution of the total dislocation length across the five-model UNEP-v1 ensemble is shown in Fig.~\ref{fig10-shock-config}(b). 
All models exhibit a consistent trend: a sharp increase in dislocation density occurs before 10 ps, followed by a plateau between 10--18 ps. 
A rapid decrease then follows, nearly vanishing by 26 ps. 
Beyond 30 ps, the dislocation length increases once more, peaking around 34 ps before declining again. 
Despite this qualitative agreement, significant quantitative deviations are observed between the models, particularly during the 10--18 ps plateau, where the maximum dislocation length varies considerably. 
This variability in microstructural evolution underlies the spatial deviations in temperature and density observed in the spall region in Fig.~\ref{fig9-shock-analysis}(d--f), highlighting that while the global spall strength is robust, the local damage microstructure is more model-sensitive. 

We further note that, although UNEP-v1 exhibits severe energy drift at 3000 K in the long-time NVE stability tests, the local temperature reached during the present shock-spallation simulation remains below 2000 K. 
As shown in Fig.~\ref{fig9-shock-analysis}(c), the maximum local temperature near the shock front and spall region lies within the temperature range where UNEP-v1 does not show catastrophic energy drift in Fig.~\ref{fig5-hea}(a). 
In addition, the shock-loading stage lasts only 10 ps, much shorter than the 100 ps NVE stability test, which further reduces the likelihood that the long-time high-temperature drift observed at 3000 K dominates the present shock response. 
Nevertheless, the use of NEP under more extreme shock conditions should be treated with caution, especially if the local temperature approaches the range where high-temperature energy drift becomes significant. 
Beyond this stability consideration, the ensemble results also reveal a distinction between global and local uncertainty. 
Since the UNEP-v1 models used here are taken from the published ensemble, the present simulations do not allow us to rigorously separate parametric uncertainty arising from different model initializations or hyperparameters from epistemic uncertainty associated with limited training coverage of high-strain-rate nonequilibrium environments. 
However, given that the original UNEP-v1 training dataset was not specifically optimized for shock loading, the observed model-to-model variations in local density, temperature, and dislocation evolution likely indicate larger uncertainty in local microstructural pathways than in the global spall strength. 
The small ensemble variance in spall strength should therefore not be interpreted as equally small uncertainty in all local atomistic fields. 
We also note that a comparable GRACE-FS shock simulation at the same scale is computationally impractical in the present work, while a substantially reduced-size GRACE-FS simulation would not provide a direct validation of spallation behavior because shock-wave propagation, void nucleation, and damage evolution are strongly affected by simulation size and geometry.

\section{\label{discussion}Discussion}

MLIPs enable near-DFT accuracy at orders-of-magnitude lower computational cost, which is particularly valuable for compositionally complex materials where multicomponent chemistry governs defect formation, deformation, and failure. In this work, we performed a controlled comparison of UNEP-v1 and GRACE-FS for 16 elemental metals and their multicomponent alloys, focusing on the practical requirements for large-scale MD simulations: accuracy, computational efficiency, finite-temperature stability, chemical transferability, and uncertainty quantification.

Our results reveal a clear trade-off between inference speed and predictive robustness. UNEP-v1 exhibits a substantial advantage in computational throughput, reaching approximately 40 times higher inference speed than GRACE-FS-M in large-scale MD simulations. This makes UNEP-v1 particularly attractive when system size, simulation timescale, or sampling efficiency is the primary bottleneck. By contrast, GRACE-FS-M shows higher training efficiency, better average energy and force accuracy, stronger chemical transferability, and more stable energy conservation in high-temperature MD simulations. Therefore, rather than identifying a single universally superior model, the present results suggest that the two frameworks occupy complementary positions. GRACE-FS-M is preferable when accuracy, chemical extrapolation, and finite-temperature robustness are the main requirements, whereas UNEP-v1 is preferable when million-atom throughput and long-time MD sampling are the limiting factors.

These findings also have important implications for developing MLIPs for multicomponent systems, including high-entropy alloys, high-entropy ceramics, and compositionally complex two-dimensional materials. Previous work suggested that training on unary and binary structures can already provide a reasonable description of high-entropy alloys~\cite{song2024-NEP16}. Our results partly support this view: the unary-plus-binary strategy is an efficient starting point because it greatly reduces the combinatorial chemical space while still yielding nontrivial accuracy for multicomponent environments. However, the systematic increase in prediction error for structures containing three or more elements indicates that unary and binary data alone are generally insufficient to guarantee robust transferability in truly high-order chemical environments. Therefore, although unary and binary datasets provide a useful foundation, incorporating ternary and higher-order configurations remains important for improving reliability in realistic multicomponent applications.

Importantly, the amount of higher-order training data required is architecture-dependent. The chemical transferability benchmark shows a clear hierarchy, with GRACE-2L outperforming GRACE-FS-M, and GRACE-FS-M outperforming UNEP-v1. This trend suggests that model architecture can partly reduce, although not eliminate, the need for exhaustive multicomponent training data. More expressive architectures may share chemical information more effectively across elemental combinations, thereby improving extrapolation from unary and binary environments to higher-order alloys. This conclusion is consistent with recent progress in universal MLIPs, such as MACE and eqV2, where expressive equivariant architectures have enabled high accuracy across broad compositional spaces, in some cases even without system-specific fine-tuning~\cite{Shuang2025-uMLIP-benchmark}. At the same time, the present results emphasize that broad elemental coverage alone is not sufficient: finite-temperature stability, chemical extrapolation, and uncertainty behavior must also be tested explicitly before deploying MLIPs in demanding MD simulations.

Finally, it is important to emphasize the distinct advantage of the NEP framework. Although the training of UNEP-v1 is relatively slow, a limitation that future gradient-based training algorithms may help address~\cite{gNEP}, its inference efficiency in GPUMD is exceptional. This efficiency provides substantial practical flexibility: one can simulate larger systems, access longer timescales, perform broader ensemble sampling, or use more physically relevant strain rates than would be feasible with more computationally expensive MLIPs. In the present work, this capability enables three-million-atom shock simulations of a prototypical FCC high-entropy alloy. The ensemble results show a small spread in the predicted spall strength, supporting the robustness of this global observable under the present loading condition. Nevertheless, the ensemble variations in local density, temperature, and dislocation evolution indicate that local atomistic pathways under shock loading remain more uncertain than the global spall strength. Thus, while NEP's exceptional speed enables million-atom shock simulations that are currently impractical with GRACE-FS, future studies focused on local shock-induced mechanisms may require training data specifically enriched with high-strain-rate nonequilibrium configurations.

Overall, this study highlights that deploying MLIPs for multicomponent alloys is not only a question of achieving low static test errors. Reliable large-scale simulations require a balanced assessment of accuracy, speed, chemical transferability, finite-temperature stability, and uncertainty quantification. In this sense, GRACE-FS and UNEP-v1 provide complementary routes: GRACE-FS offers stronger robustness and transferability, while UNEP-v1 offers exceptional throughput for extreme-scale MD. Future MLIP development for compositionally complex materials should therefore combine architecture-aware chemical representation, targeted higher-order training data, systematic finite-temperature stability tests, and ensemble-based uncertainty quantification.

\section{\label{conclusion}Conclusion}

In summary, this work presents a controlled benchmark of the UNEP-v1 and GRACE-FS MLIP frameworks for 16 elemental metals and their multicomponent alloys. The results reveal a clear accuracy--efficiency trade-off. GRACE-FS offers higher training efficiency, better average accuracy, stronger chemical transferability, and improved finite-temperature stability, whereas NEP provides exceptional inference speed and therefore enables molecular dynamics simulations at the million-atom scale. We further show that chemical transferability is closely connected to high-temperature MD stability in highly multicomponent environments, highlighting the importance of architecture-dependent chemical representation for reliable simulations of complex alloys. For uncertainty quantification, ensemble-based uncertainty provides a more reliable error indicator than D-optimality in the present heterogeneous dataset. Finally, the NEP ensemble enables three-million-atom shock-spallation simulations of a high-entropy alloy, yielding robust global spall-strength predictions while also revealing larger uncertainty in local damage pathways. These findings demonstrate that reliable deployment of MLIPs for multicomponent alloys requires the joint consideration of accuracy, speed, transferability, stability, and uncertainty, and provide practical guidance for selecting MLIP frameworks for large-scale simulations under extreme conditions.

\section{\label{data}Data availability}
All the GRACE-FS, GRACE-2L models and validation DFT dataset for multicomponnenet alloys are available at https://doi.org/10.5281/zenodo.19187064.

\section{\label{data}Acknowledgments}
This work was sponsored by Nederlandse Organisatie voor WetenschappelijkOnderzoek (The Netherlands Organization for Scientific Research, NWO) domain Science for the use of supercomputer facilities. The authors also acknowledge the use of DelftBlue supercomputer, provided by Delft High Performance Computing Center (https://www.tudelft.nl/dhpc). This work was supported by the National Outstanding Youth Science Fund Project (number 12125206), Major International Joint Research Projects (number W2411003) of NSFC.

\section{Appendix}

\begin{table*}[ht]
    \setlength{\tabcolsep}{7pt}
    \centering
    \caption{Ensemble-averaged mean absolute errors (MAE) and root-mean-square errors (RMSE) of energy ($E$), force ($F$), and stress ($S$) for UNEP-v1 and GRACE-FS-M on the training and test sets. Values are reported as mean $\pm$ standard deviation over six ensemble models. The stress errors are converted to GPa by multiplying by 160.2177.}
    \label{tab-ensemble-error-uncertainty}
    \begin{tabular}{lcccc}
        \toprule
        \multirow{2}{*}{Metric} 
        & \multicolumn{2}{c}{Training set} 
        & \multicolumn{2}{c}{Test set} \\
        \cmidrule(lr){2-3} \cmidrule(lr){4-5}
        & UNEP-v1 & GRACE-FS-M & UNEP-v1 & GRACE-FS-M \\
        \midrule
        $E_{\mathrm{MAE}}$ (meV/atom) 
        & $10.69 \pm 0.58$ 
        & $10.38 \pm 0.15$ 
        & $23.82 \pm 1.56$ 
        & $20.66 \pm 1.02$ \\
        
        $E_{\mathrm{RMSE}}$ (meV/atom) 
        & $16.49 \pm 0.97$ 
        & $20.77 \pm 0.61$ 
        & $54.99 \pm 3.74$ 
        & $73.90 \pm 13.71$ \\
        
        $F_{\mathrm{MAE}}$ (meV/\AA) 
        & $109.27 \pm 2.23$ 
        & $99.76 \pm 0.40$ 
        & $115.27 \pm 2.30$ 
        & $99.63 \pm 0.91$ \\
        
        $F_{\mathrm{RMSE}}$ (meV/\AA) 
        & $169.11 \pm 3.56$ 
        & $160.27 \pm 0.70$ 
        & $271.24 \pm 9.72$ 
        & $469.19 \pm 37.52$ \\
        
        $S_{\mathrm{MAE}}$ (GPa) 
        & $0.44 \pm 0.02$ 
        & $0.52 \pm 0.01$ 
        & $0.68 \pm 0.04$ 
        & $0.65 \pm 0.02$ \\
        
        $S_{\mathrm{RMSE}}$ (GPa) 
        & $1.20 \pm 0.09$ 
        & $2.34 \pm 0.12$ 
        & $2.16 \pm 0.22$ 
        & $2.99 \pm 0.41$ \\
        \bottomrule
    \end{tabular}
\end{table*}


\bibliography{apssamp}

\end{document}